\documentclass[12pt,preprint]{aastex}

\newcommand{\Mearth}{M$_\oplus$}
\newcommand{\Msun}{M$_\odot$}

\newcommand{\Rearth}{R$_\oplus$}

\newcommand{\Lsun}{L$_\odot$}
\newcommand{\AU}{{\sc au}}

\slugcomment{}

\shorttitle{Is OGLE-2005-BLG-390Lb entirely frozen?}

\shortauthors{Ehrenreich et al.}

\begin{document}

\title{On the Possible Properties of Small and Cold Extrasolar Planets:\\
Is OGLE-2005-BLG-390Lb Entirely Frozen?}

\author{David Ehrenreich, Alain Lecavelier des Etangs, and Jean-Philippe Beaulieu}
\affil{Institut d'astrophysique de Paris \\
    CNRS (UMR 7095) ; Universit\'e Pierre \& Marie Curie \\
    98 bis, boulevard Arago 75014 Paris, France}
\email{ehrenreich@iap.fr}

\and

\author{Olivier Grasset}
\affil{Laboratoire de plan\'etologie et de g\'eodynamique \\
    CNRS (UMR 6112) ; Universit\'e de Nantes \\
    2, rue de la Houssini\`ere 44322 Nantes, France}

\begin{abstract}
Extrasolar planets as light as a few Earths are now being detected. Such planets are likely not
gas or ice giants. Here, we present a study on the possible properties of the small and cold
extrasolar planets, applied to the case of the recently discovered planet OGLE-2005-BLG-390Lb
(Beaulieu et al.\ 2006). This planet ($5.5^{+5.5}_{-2.7}$ Earth masses) orbits
$2.6^{+1.5}_{-0.6}$-astronomical units away from an old M-type star of the Galactic Bulge. The
planet should be entirely frozen given the low surface temperature (35 to 47~K). However,
depending on the rock-to-ice mass ratio in the planet, the radiogenic heating could be sufficient
to make the existence of liquid water within an icy crust possible. This possibility is estimated
as a function of the planetary mass and the illumination received from the parent star, both being
strongly related by the observational constraints. The results are presented for water-poor and
water-rich planets. We find that no oceans can be present in any cases at 9--10~Gyr, a typical age
for a star of the Bulge. However, we find that, in the past when the planet was $\la5$~billion
years old, liquid water was likely present below an icy surface. Nevertheless, the planet is now
likely to be entirely frozen.
\end{abstract}


\section{INTRODUCTION}

Some 190 extrasolar planets have been detected since 1992.\footnote{From
J.~Schneider's \emph{Extrasolar planets encyclod\ae dia} \\ at
\texttt{http://vo.obspm.fr/exoplanetes/encyclo/encycl.html}.} Most of these
discoveries have been performed using radial velocimetry. Recent progresses and
new instruments have permitted the lowering of the mass detection threshold to
tens of Earth masses (\Mearth). For instance, a 14-\Mearth\ ($\times \sin{i}$,
where $i$ is the inclination of the planet orbital plane) planet around the
star \object{$\mu$~Arae} (Santos et al.\ 2004), a 7.5-\Mearth\ planet around
\object{GJ~876} (Rivera et al.\ 2005), and a 10-\Mearth\ planet around
\object{HD~69\,830} (Lovis et al.\ 2006) have been reported. All these objects
lie at the mass boundary between telluric and ice giant planets, as it is
inferred from analogy with the Solar System ice giants, \object{Uranus} and
\object{Neptune} (14.5 and 17.2~\Mearth, respectively). Besides, all low-mass
planets discovered with radial velocimetry are orbiting very close to their
parent stars, $\la 0.1$~astronomical units (\AU), so they are enduring extreme
radiative heating (Lecavelier des Etangs 2006), an unknown situation in the
Solar System. Indeed, the radial-velocity technique is biased towards
short-orbital-period planets.

On the contrary, the microlensing technique is sensible to planets with larger
semi-major axis, hence it allows to detect a different kind of planets. Using
this technique, Beaulieu et al.\ (2006) have detected a
$5.5^{+5.5}_{-2.7}$-\Mearth\ planet orbiting at $2.6^{+1.5}_{-0.6}$~\AU\ from a
$0.22^{+0.21}_{-0.11}$-solar-mass (\Msun) star. It is potentially the lightest
exoplanet detected so far. This star and its planet are the lens of the
gravitational microlensing event OGLE-2005-BLG-390 (see Table~\ref{tab:lens})
and the star, according to its mass range, is a M star, with a luminosity
reaching only $\sim0.01$~solar luminosity (\Lsun). Hence, the amount of energy
per surface unit this planet receives from its star is comparable to that of
Pluto in the Solar System ($\sim0.1$~W\,m$^{-2}$). The planet detected by
Beaulieu et al.\ (2006), poetically designated as OGLE-2005-BLG-390Lb, is thus
a cold sub-Uranus object.

In the light of the orbital elements and masses presented in Beaulieu et al.\ (2006), the planet
can be a cold and massive analog of the Earth, or alternatively be similar to a frozen
ocean-planet having retained a lot of water (L\'eger et al.\ 2004). In both cases, the question
arises if the planet is entirely frozen or if liquid water can exist close to the surface. To
answer that question, we focus on the phase study of water (H$_2$O) in the first $\sim100$~km
under the planetary surface, using observational constrains and similarities with icy satellites
of the Solar System. A full model of the entire planet is certainly beyond the scope of the
present work (see Sotin et al.\ 2006 for a detailed study of the internal structure of these kinds
of extrasolar planets).

\section{ASSUMPTIONS AND BASIC PROPERTIES OF OGLE-2005-BLG-390Lb}
\label{sec:basic_props}

All the parameters of the lens system deduced from the detection are related to the mass of the
lens star $M_\star$. The mass of the planet $M_P$ is simply given by the relation
\begin{equation}
\label{eq:Mplanet_to_Mstar} M_P = q \times M_\star,
\end{equation}
where $q$ is the planet-to-star mass ratio measured by Beaulieu et al.\ (2006) and given in
Table~\ref{tab:lens}. We took advantage of these correlations (eqs.~[\ref{eq:Mplanet_to_Mstar},
\ref{eq:poly_a}, and \ref{eq:L_star}]) to estimate the basic properties of the system as a
function of the stellar and planetary masses, such as the semi-major axis of the planet and the
stellar luminosity (\S~\ref{sec:constraints}). Then, we calculated the energy the planet is
receiving from its star and its equilibrium temperature, again as a function of the planetary mass
(\S~\ref{sec:temp}). Given the surface temperature and the atmospheric scale height, we then
discuss on the possible nature of an atmosphere (\S~\ref{sec:atmo}).

\subsection{Constraints from the Microlensing Event}
\label{sec:constraints} The microlensing event OGLE-2005-BLG-390 is described by Beaulieu et al.\
(2006), and the relevant parameters for this study are presented in Table~\ref{tab:lens}. All the
parameters are in fact represented by a Bayesian distribution of probability. We have explored the
phase space allowed by the results presented in the above-cited paper.

Beaulieu et al.'s analysis (based on Dominik 2006) yields 95\% probability that
the lens star be a main sequence star, 4\% that it be a white dwarf, and less
than 1\% that it be a neutron star or a black hole. They find a 75\%
probability that the star be part of the Galactic Bulge. When adopting a source
star distance of 8.5~kpc, the lens star is estimated to be $6.6\pm1.0$-kpc
away, toward the Galactic Center. Zoccali et al.\ (2003) have shown that the
Bulge is about 10-billion-years (Gyr) old with no evidence for younger stellar
populations. They also proved that the Bulge metallicity is peaked near the
solar value, with a sharp cutoff just above this value and a tail towards lower
metallicity. As part of the Bulge population, we therefore consider that the
lens system (star and planet) is likely to be $\sim10$-Gyr old, with a solar
metallicity.

The mass of the stellar lens derived from Beaulieu et al.\ is $M_\star \approx 0.22$~\Msun, within
a range [0.11,0.43]~\Msun,\footnote{The quantity accurately measured by Beaulieu et al.\ (2006) is
the planet-to-star mass ratio $q$.} i.e., most probably the mass of a M-type star. In that range
and using eq.~(\ref{eq:Mplanet_to_Mstar}), the planetary mass is found within [3,11]~\Mearth; the
planet has more than 95\% probability to be lighter than \object{Uranus}.

The separation of the lens $d$ is expressed in Einstein's radii, $R_\mathrm{E}$. From the equation
defining $R_\mathrm{E}$ as a function of $M_\star$ (see Table~1 in Beaulieu et al.\ 2006), the
semi-major axis $a$ of the planet can be expressed as
\begin{equation}
    \label{eq:poly_a} \frac{a}{1~\textrm{\AU}} = 0.53 + 9.76 \frac{M_\star}{M_\odot}
    - 6.30 \left(\frac{M_\star}{M_\odot}\right)^2
    + 2.59 \left(\frac{M_\star}{M_\odot}\right)^3.
\end{equation}
Given the stellar-mass range explored, $a$ is within [1.5,3.8]~\AU.

To estimate the surface temperature of OGLE-2005-BLG-390Lb (see
\S~\ref{sec:temp}) we need first estimate the stellar luminosity $L_\star$.
This is done using a mass-luminosity relation derived from Baraffe et al.\
(1998) for a solar metallicity, (see also Beaulieu et al. 2006b, in
preparation)
\begin{eqnarray}
\label{eq:L_star}
        \log_{10}{\frac{L_\star}{L_\odot}} &&= -0.04 + 6.03 \log_{10}{\frac{M_\star}{M_\odot}}
        + 4.16 \left(\log_{10}{\frac{M_\star}{M_\odot}}\right)^2 \nonumber\\
        &&-2.19 \left(\log_{10}{\frac{M_\star}{M_\odot}}\right)^3
        - 3.37 \left(\log_{10}{\frac{M_\star}{M_\odot}}\right)^4.
\end{eqnarray}
Considering again the same stellar-mass range than above, the luminosity of the
planet host star should be within [0.001,0.02]~\Lsun. This result would be
barely changed by assuming a different metallicity than solar; for instance, a
0.5-dex decrease in metallicity would increase the luminosity by $\sim20\%$
(still according to Baraffe et al.\ 1998), then triggering only a $\sim1$-K
increase in the surface temperature of the planet (see below).

\subsection{Energy Received from the Star}
\label{sec:temp} From the semi-major axis of the planet and the stellar
luminosity, it is straightforward to calculate the amount of energy received by
the planet and thus its equilibrium surface temperature, which also depends on
the surface albedo. The power received per surface unit is $q_\star =
L_\star/\left(16\pi a^2\right)$, i.e., within [0.2,0.8]~W\,m$^{-2}$. This is
roughly the same amount of energy received by Pluto or Neptune from the Sun.
Analogously to the outer planets of the Solar System, which formed and remained
beyond the snowline, OGLE-2005-BLG-390Lb can contain an important fraction of
volatile material (ices) and thus ressemble a massive Ganymede with frozen
surface.

\begin{figure}
\resizebox{\hsize}{!}{\includegraphics{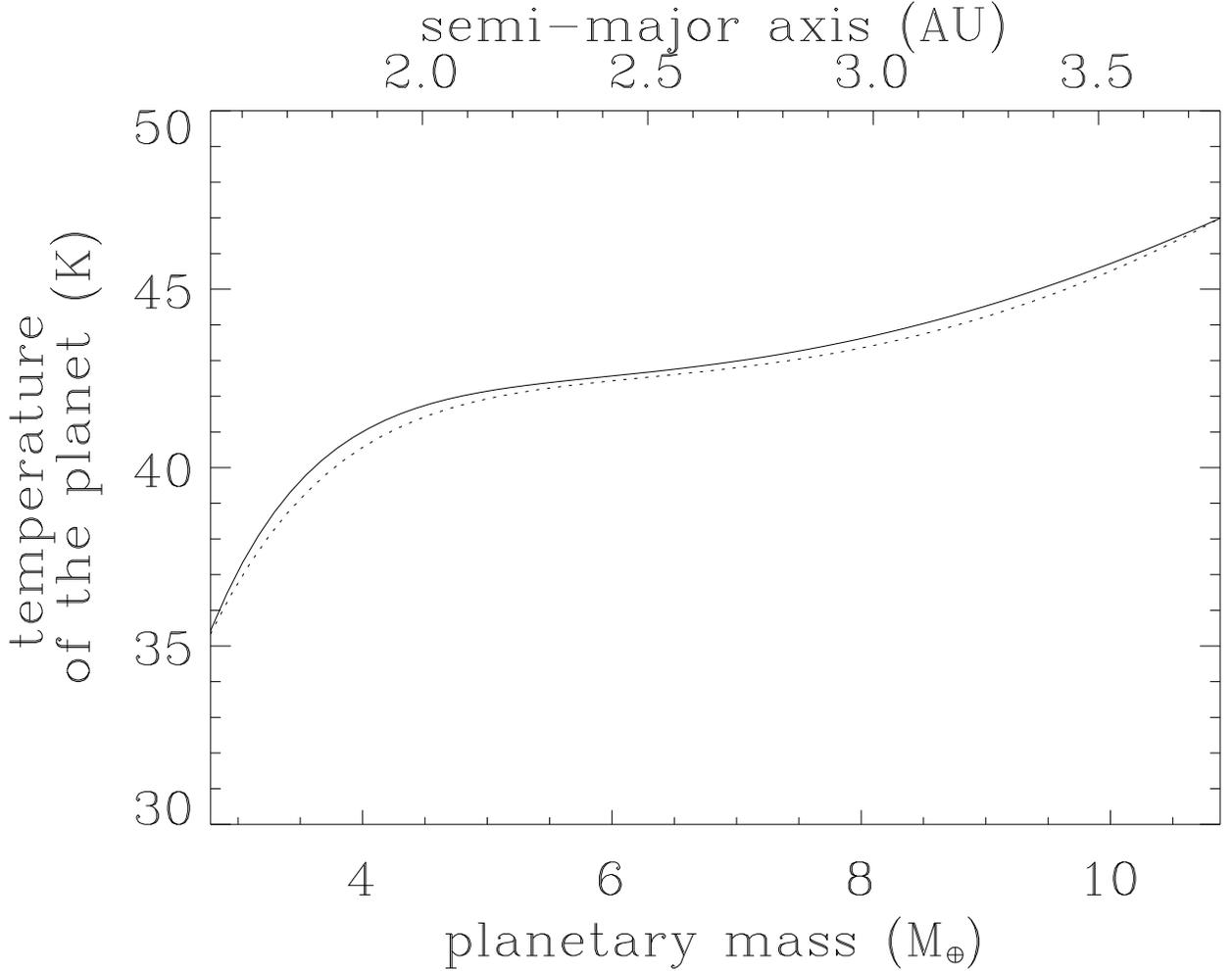}} \caption{Possible surface
temperatures of the planet. In our case, the properties of the star are
depending on the mass of the planet ($M_\star = M_P / q$ where $q =
7.6\times10^{-5}$, and $L_\star \equiv L_\star(M_\star)$, as expressed in
eq.~[\ref{eq:Mplanet_to_Mstar}] and eq.~[\ref{eq:L_star}], respectively). In
addition, the semi-major axis depends on $M_\star$ (eq.~[\ref{eq:poly_a}]) and
thus on $M_P$. The surface temperature is calculated assuming it is equal to
the equilibrium temperature with an albedo of 50\%, and is represented here as
a function of the planetary mass ($T_\mathrm{surf} \equiv
T_\mathrm{surf}(M_P)$, bottom axis, plain line) and semi-major axis
($T_\mathrm{surf} \equiv T_\mathrm{surf}\left(a\left[M_P\right]\right)$, top
axis, dotted line), accordingly to eq.~(\ref{eq:T_eq}). As a result, the
temperature of the planet evolves counter-intuitively with the semi-major axis:
the larger the separation between the planet and its star, the \emph{higher}
the surface temperature is. This is because the stellar luminosity is not
constant, but rather becomes more intense as both the planetary and stellar
masses increase (eq.~[\ref{eq:L_star}]). \label{fig:T_vs_M}}
\end{figure}

We then assume an albedo $A$ of 50\% for an ice-covered surface and calculate the surface
temperature of the planet, taken to be the equilibrium temperature,
\begin{equation}
\label{eq:T_eq} T_\mathrm{surf} \approx T_\mathrm{eq} = \sqrt[4]{\frac{L_\star
\left(M_\star\right) \left(1-A\right)}{16 \pi a^2 \sigma}},
\end{equation}
where $\sigma$ is the Stefan-Boltzmann constant and $L_\star(M_\star)$ is the
stellar luminosity as a function of the stellar mass, as given by
eq.~(\ref{eq:L_star}). As $M_\star$ is related to $M_P$ through
eq.~(\ref{eq:Mplanet_to_Mstar}), the stellar luminosity is also a function of
$M_P$. Figure~\ref{fig:T_vs_M} represents the surface temperature as a function
of the planetary mass. Considering the parameter range given above,
$T_\mathrm{surf}$ is within [35,47]~K.

\subsection{Atmosphere}
\label{sec:atmo} Given the surface temperature range, the vapor pressure of most volatile species
of interest (H$_2$O, methane CH$_4$, ammonia NH$_3$, carbon dioxide CO$_2$) is very weak, below
0.01~Pa. Only carbon monoxide (CO) and molecular nitrogen (N$_2$) have relatively high vapor
pressures at these temperatures ($p_{\mathrm{CO}_2} \approx 100$~Pa at 47~K, $p_{\mathrm{N}_2}
\approx 0.1$~Pa at 35~K). Moreover, N$_2$  starts sublimating around 47~K, where $p_{\mathrm{N}_2}
\approx 100$~Pa.\footnote{Vapor pressure data can be found for various species on Air Liquide's
Gas Encyclopaedia website at \texttt{http://encyclopedia.airliquide.com} while temperatures for
different vapor pressures may be found on AVS Technical Resources' website at
\texttt{http://www.aip.org/avsguide/refguide/vapor.html}.} Thus, these two species are the most
likely to be present in the planetary atmosphere.

For a planetary mass within [3,11]~\Mearth, and given the radii we estimate in
\S~\ref{sec:radius}, the surface gravity of OGLE-2005-BLG-390Lb lies within
[10,30]~m\,s$^{-2}$. We consequently estimate the atmospheric scale height for
N$_2$ or CO to be within [0.5,1]~km. Considering the small scale height and
vapor pressure for N$_2$, we hypothesize the presence of a tenuous atmosphere
of N$_2$ with a surface pressure $p_0 = p_{\mathrm{N}_2} \sim 1$~hPa. Note that
the atmospheric pressure measured at the surface of Pluto (0.5~Pa according to
Sicardy et al.\ 2003) is lower than the vapor pressure of N$_2$ at Pluto's
surface temperature ($\sim 40$~K), therefore, the nitrogen vapor pressure of
1~hPa can be considered as an upper limit to the actual atmospheric pressure in
our case. Anyway, it is a fairly weak value, insignificant for the calculation
of the ice shell structure (see \S~\ref{sec:liquid_now}). We further take an
atmospheric pressure at the surface $p_0 \approx 0$.

The main properties of OGLE-2005-BLG-390Lb deduced from the measurements of Beaulieu et al.\
(2006) suggest an extremely cold surface. If liquid water is to exist somewhere in this planet, it
is very likely below an icy crust, as in the icy satellites of the Solar System.

\section{STRUCTURE OF THE PLANET}
\label{sec:structure} We saw in the previous section that liquid water cannot exist on the surface
of OGLE-2005-BLG-390Lb because of the low temperature and pressure. Instead, the surface of the
planet should be frozen ice. The question is then: can liquid water exist below the ice layer? A
description of the physical states of water in the ice shell can be obtained using a phase diagram
and knowing the evolution of the pressure and temperature within the ice shell. The pressure at a
depth $z$ in the ice shell is
\begin{equation}
\label{eq:p(z)} p(z) = \int_{0}^{z} \rho(x) g(x) \mathrm{d}x;
\end{equation}
it depends on the gravity of the planet, hence on its radius.

\subsection{Internal Structure and Radius of OGLE-2005-BLG-390Lb}
\label{sec:radius} Models of possible internal structures for Earth-mass
extrasolar planets have already been developed. The radius of
OGLE-2005-BLG-390Lb is found using such a model (Sotin et al.\ 2006), which a
basic version has been used by L\'eger et al.\ (2004) for the modeling of
`ocean-planets'. Its physical approach is analogous to that of Valencia et al.\
(2006). The main difference is that the bulk composition of the planet is fixed
by the composition of the star. The input parameters are the stellar
iron-to-silicium [Fe/Si] and magnesium-to-silicium [Mg/Si] ratios, the
magnesium content of the silicate mantle $\mathrm{Mg_\#} = [\mathrm{Mg} /
(\mathrm{Mg + Fe})]$, the amount of H$_2$O, and the total mass of the planet
(Sotin et al.\ 2006).

Solar [Fe/Si] and [Mg/Si] are assumed together with a terrestrial Mg$_\# =
0.9$. It can be noted that varying these numbers from the solar and terrestrial
values changes the size of the metallic core (composed by Fe and iron sulphide
FeS) and the amount of Fe in the silicate mantle, but causes small differences
on the computed planetary radii (3/1000 in average). On the contrary, the
amount of H$_2$O strongly influences the radius for a given planetary mass. We
consider different enrichments of the planet in H$_2$O, namely 0.025, 25, and
50\% of the planetary mass.\footnote{In the following, fractions of water will
be given as fractions of the planetary mass and noted as weight-percentages
(\%wt).}

Mass-radius relationships are then derived using appropriate equations of state
for the different layers. The Mie-Gr\"uneisen-Debye formulation has been chosen
for describing the iron core (Uchida et al.\ 2001), the lower silicate mantle
(Hemley et al.\ 1992) and the icy mantle (Fei et al.\ 1993). The
3$^\mathrm{rd}$-order Birch-M\"urnaghan equation of state is used for the upper
silicate mantle (Vacher et al.\ 1998) and for the liquid layer (Lide 2005).

The 0.025\%wt-H$_2$O~case corresponds to a rocky planet with a H$_2$O content
similar to the Earth one. The model yields a structure consisting of a metallic
core accounting for 32\%wt and a rocky mantle (silicates) accounting for
$f_\mathrm{sil} = 68\%$wt, where $f_\mathrm{sil}$ represents the mass fraction
of silicates in the planet. The resulting radius is within [1.3,2.0]~\Rearth\
depending on the planetary mass.

\begin{figure}
\resizebox{\hsize}{!}{\includegraphics{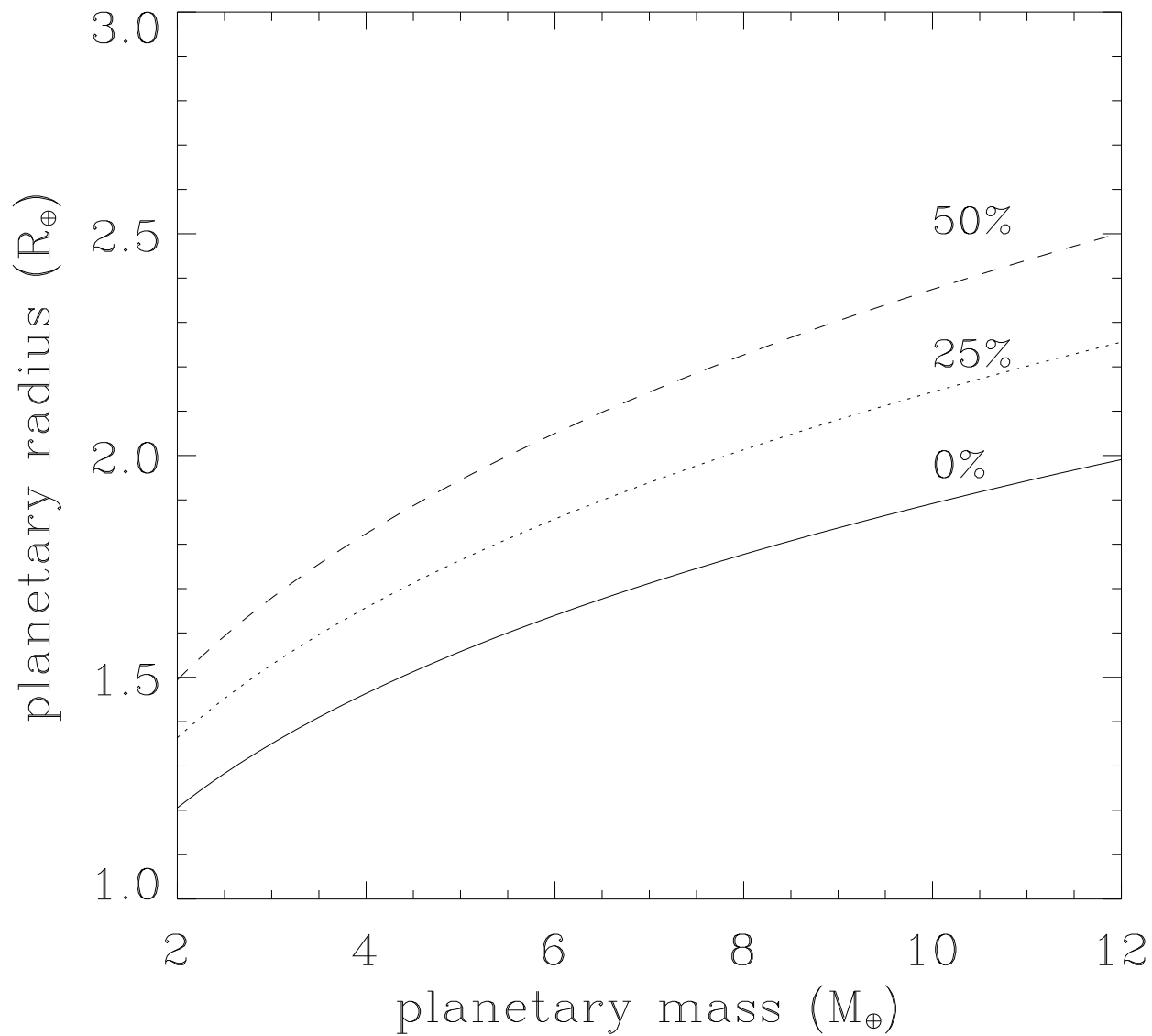}} \caption{Radius of the planet as a function of the
planetary mass. The radius is calculated using a mass-radius relation from Sotin et al.\ (2006),
for different H$_2$O mass fractions: 0.025\%wt (noted 0\%wt, plain line), 25\%wt (dotted line),
and 50\%wt (dashed line). \label{fig:planetprop}}
\end{figure}

For both other cases, 25 and 50\%wt-H$_2$O, the silicate\--mantle\--to\--iron\--core mass ratio
remains similar to the 0.025\%wt case, i.e., about 2/1. Over the whole planet, the silicates must
account for $f_\mathrm{sil} = 0.68 \times (1 - 0.25) = 51\%$wt and $f_\mathrm{sil} = 0.68 \times
(1 - 0.5) = 34\%$wt in the 25 and 50\%wt cases, respectively. Planetary radii are found wihtin
[1.4,2.3]~\Rearth\ and [1.6,2.5]~\Rearth, respectively, depending on the planetary mass. These
results are summerized in Fig.~\ref{fig:planetprop}.


\subsection{Phases of Ice in the Icy Mantle}
\label{sec:hp_ices} Due to the low surface temperature, the volatile material
present in OGLE-2005-BLG-390Lb must be condensed as an ice shell. Since the
planet is massive enough to be completely differentiated, the ice shell must be
overtopping a denser rocky core, in a situation similar to that of the icy
moons of giant planets in the Solar System.

\subsubsection{The Water-Poor Case (0.025\%wt)}
\label{sec:water-poor shell} 
\begin{figure}
\resizebox{\hsize}{!}{\includegraphics{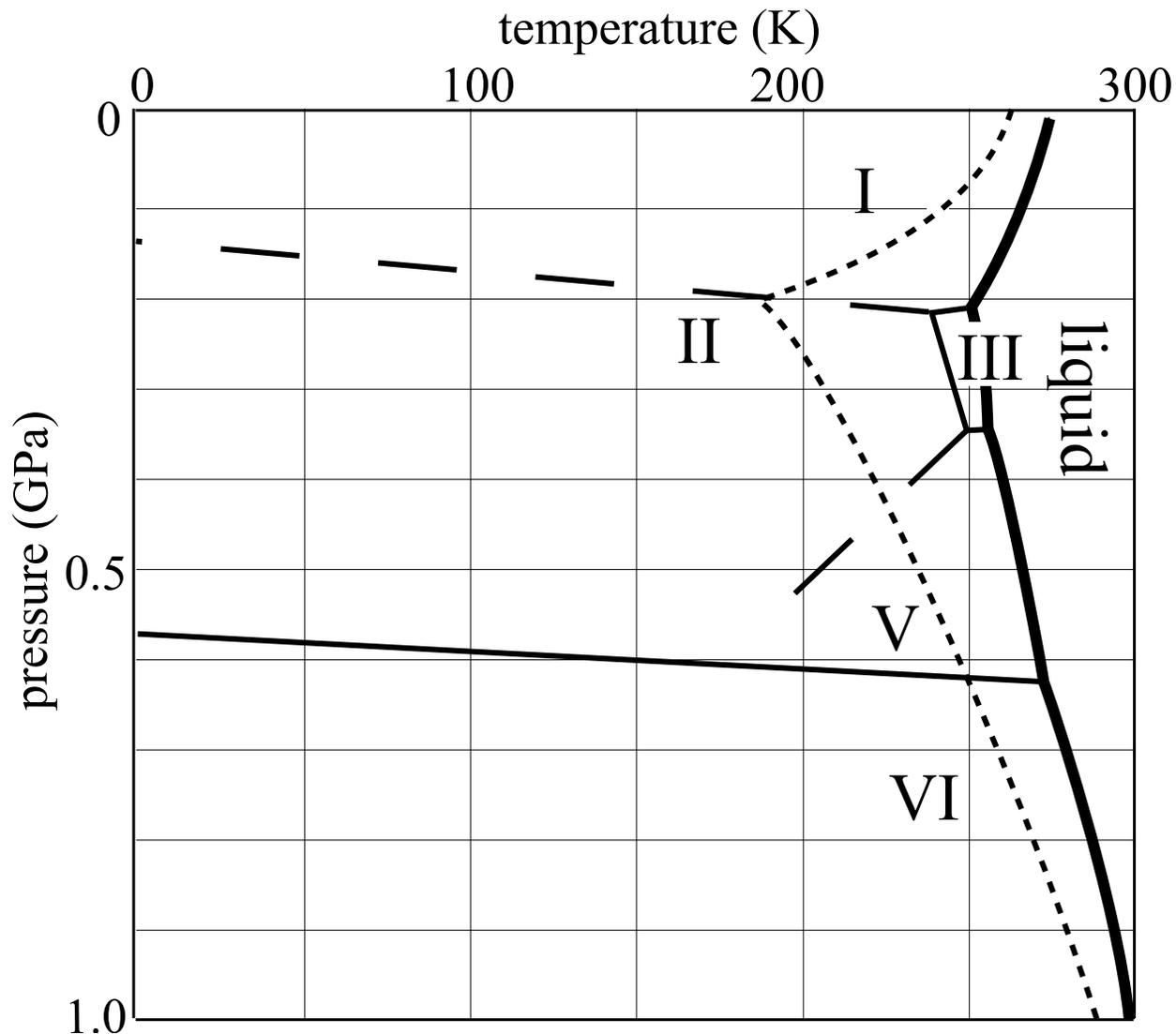}} \caption{Phase diagram of water. The
domains of existence in pressure and temperature of the liquid phase and five solid phases are
figured. The thick line is the phase boundary between the different ices and liquid water, i.e.,
the melting curve. Each phase of ice corresponds to a given crystalline structure. Hexagonal
ice~{\sc i} presents the particularity of being less dense than liquid water. Besides, its melting
curve has a negative slope: ice~{\sc i} can melt under pressure if $T>250$~K. On the contrary, the
high-pressure phases represented here (ices~{\sc ii}, {\sc iii}, {\sc v}, and {\sc vi}) are all
denser than liquid water. As a consequence, a liquid layer of water can exist below an ice-{\sc i}
layer and above a ice-{\sc iii, -v,} or {\sc -vi} layer. This is the favored structure for the ice
shells of the icy moons in the Solar System, although such ice shells are usually modeled assuming
ammoniated (with NH$_3$) rather than pure liquid H$_2$O. Such a mixture crystallizes at lower
temperature than pure H$_2$O (dashed line). \label{fig:phase_diagram}}
\end{figure}

\begin{figure}
\resizebox{\hsize}{!}{\includegraphics{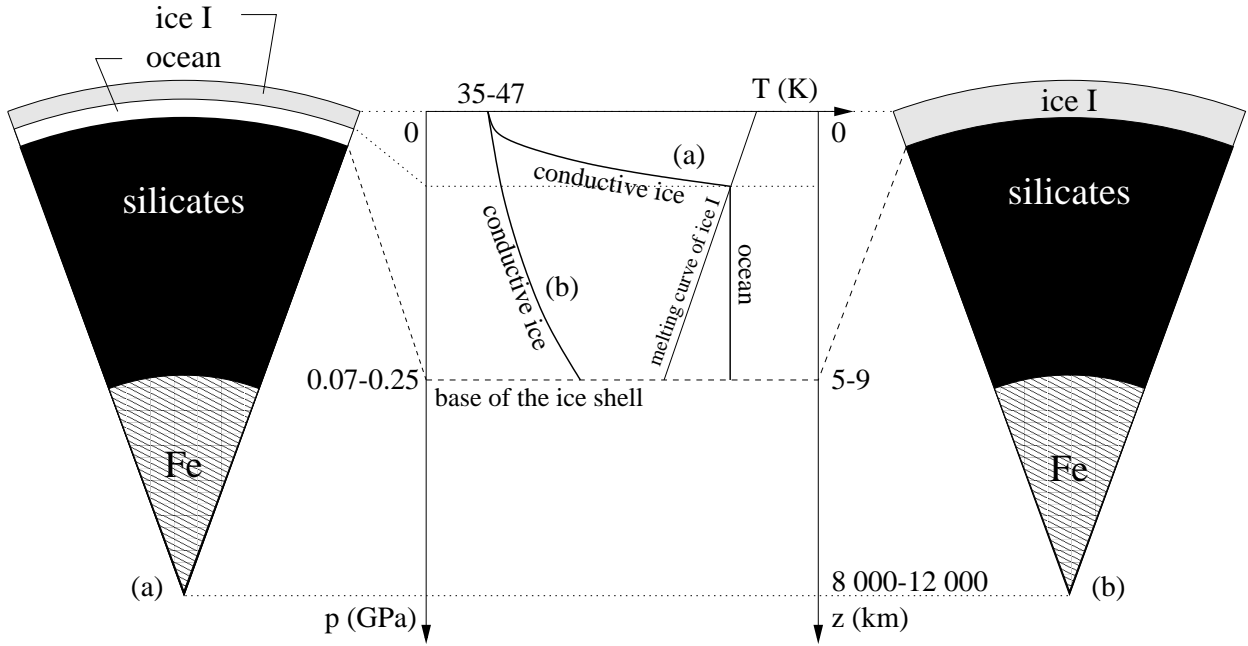}} \caption{Possible internal structures for a
water-poor OGLE-2005-BLG-390Lb, with typically a terrestrial fraction of water (0.025\%wt). In
this case, the planet is composed by a metallic core (hatching) and a silicate mantle (black). The
water can be present in the form of a thin (5--9~km) layer of low-pressure ice~{\sc i} (grey). Two
possible pressure-temperature profiles across the ice shell are sketched at the center (thick
lines), assuming the ice is conducting the heat from the base of the shell up to the surface. If
the temperature at the base of the ice shell is lower than the melting temperature of ice~{\sc i}
(thin line) then the layer is completely frozen (b). However, if the temperature at the base is
above the melting temperature then an ocean (white) can exist below the ice (a). Note that in any
cases, the H$_2$O layer would remain rather thin compared to the radius of the planet. \emph{The
figure is not to scale.} \label{fig:camembert_no_h2o}}
\end{figure}

If the planet contains the same proportion of water than the Earth,
that is, very few: $2.5 \times 10^{-4} M_P$, the maximum pressure at the bottom of the ice shell
is $p_\mathrm{ice} \approx M_\mathrm{ice} g / (4 \pi R_P^2)$, where $g$ is within
[15,29]~m\,s$^{-2}$ for a planetary mass in [3,11]~\Mearth. Then, $p_\mathrm{ice} \approx 2.5
\times 10^{-4} g^2/ (4 \pi G)$, i.e., within [0.07,0.25]~GPa. Therefore, from the phase diagram of
water represented in Fig.~\ref{fig:phase_diagram},\footnote{A full phase diagram of H$_2$O is
available on the website \emph{Water structure and behavior} by M.~Chaplin, at
\texttt{http://www.lsbu.ac.uk/water/phase.html}.} the ice shell is only composed of ice~{\sc i} or
possibly a liquid layer surrounded by a thin layer of ice~{\sc i}. Using eq.~(\ref{eq:p(z)}), the
depth of the interface ice~{\sc i}/silicates or liquid/silicates is found within [5,9]~km (see
Fig.~\ref{fig:camembert_no_h2o}). Consequently, the icy mantle that covers the silicates is rather
small compared to the planetary radius; it consists of a thin ($\la10$~km) low-pressure ice-{\sc
i} layer.

\subsubsection{The Water-Rich Case (25 and 50\%wt)}
\label{sec:water-rich shell} 
\begin{figure}
\resizebox{\hsize}{!}{\includegraphics{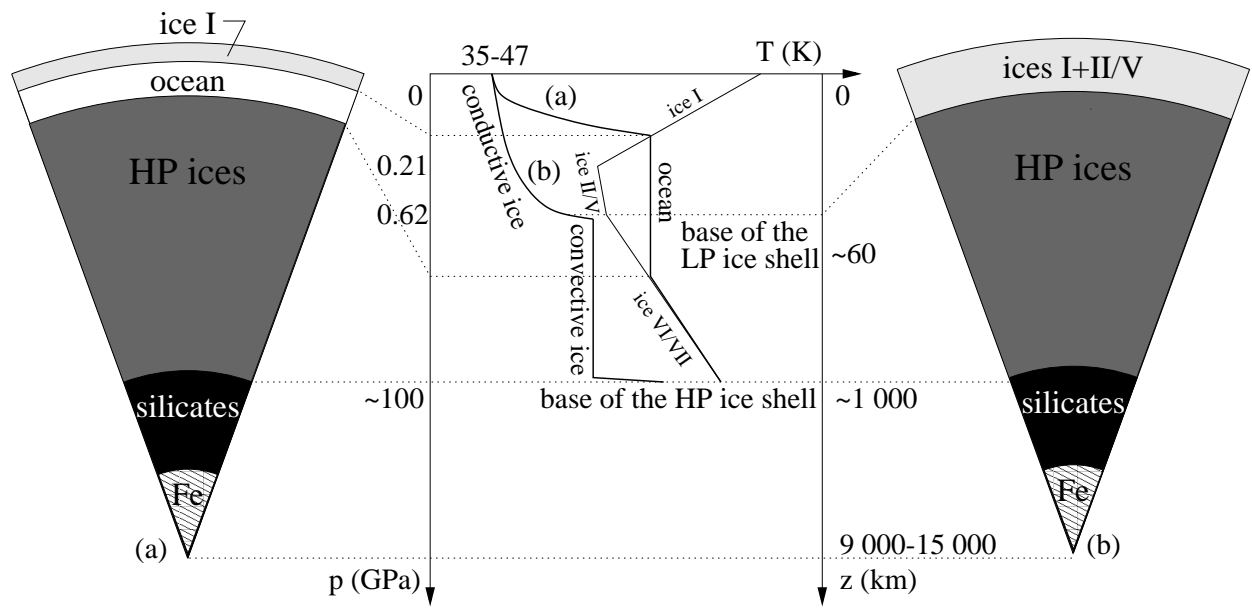}} \caption{Possible structure for a
water-rich OGLE-2005-BLG-390Lb. The mantle~(black)-to-core~(hatching) mass ratio is still fixed to
$\sim2/1$, as for the water-poor cases in Fig.~\ref{fig:camembert_no_h2o}, but here a large part
of the planetary mass is accounted for by a high-pressure ice mantle (dark grey). It is overtopped
by a low-pressure ice shell consisting in ice~{\sc i} and, depending on the temperature, liquid
water, ice~{\sc ii}, or ice~{\sc v}. The base of the low-pressure ice shell is the phase boundary
between ices~{\sc ii/v} and {\sc vi} at 0.62~GPa (see Fig.~\ref{fig:phase_diagram}). The
temperature-pressure profile across the low-pressure ice shell (light grey) is plotted at the
center in two cases (thick lines), both assuming the heat is transferred by conduction across the
low-pressure ice shell. In case (a), the profile crosses the melting curve of ice~{\sc i} (thin
line) and an ocean (white) exists between the low- and high-pressure ice shells. The temperature
profile then follows the melting curve from the base of the low-pressure ice shell down to the
base of the high-pressure ice shell. In case (b), the temperature profile does not cross the
melting curve of ice~{\sc i}, so that the low-pressure ice shell is entirely frozen. At the base
of the low-pressure ice shell, the conductive profile becomes convective. \emph{The figure is not
to scale.} \label{fig:camembert_plenty_h2o}}
\end{figure}

If a non-negligible fraction of the planetary mass is accounted for
by water, then the ice shell must be largely composed of high-pressure ice due to the strong
gravity (from 10 to 22~m\,s$^{-2}$ depending on the planetary mass and the mass percentage of
H$_2$O). The most stable phase at low temperature (10 to 200~K) and moderate pressure (0.2 to
0.6~GPa) is ice~{\sc ii}; at higher pressures (deeper in the planet), we successively meet
ice~{\sc vi}, {\sc vii}, and {\sc x} (those last two phases lie beyond the pressure scale in
Fig.~\ref{fig:phase_diagram}, at pressure greater than 2~GPa). Indeed, we find $p_\mathrm{ice}$
within [40,210]~GPa, well in the domain of high-pressure ices. In such a case, the icy mantle must
be split into an upper thin low-pressure ice-{\sc i/ii} layer and a thick high-pressure layer
composed by ices~{\sc v, vi, vii}, and {\sc x}. Actually, there is an important endothermic
frontier at the phase transitions ice~{\sc ii}/{\sc v} and {\sc ii}/{\sc vi}, which is located
around $p_{\textrm{{\sc ii}}} = 0.62$~GPa (Bercovici et al.\ 1986). With eq.~(\ref{eq:p(z)}), we
estimated the ice-{\sc i/ii} layer is never thicker than about 60~km
(Fig.~\ref{fig:camembert_plenty_h2o}). Hence, in this case we distinguish between a thin
($\la60$~km) low-pressure ice ({\sc i/ii}) and a thick ($\sim1\,000$~km) high-pressure ice ({\sc
vi/vii}) mantle.

In any cases, an entirely-solid ice shell is overtopped by a layer of low-pressure ice~{\sc i}. If
there is not much ice in the planet (\S~\ref{sec:water-poor shell}), the whole ice shell is
ice~{\sc i}. On the contrary, if the amount of ice in the planet is large (\S~\ref{sec:water-rich
shell}), high-pressure phases of H$_2$O must be found below the ice-{\sc i} lid. Because of the
negative slope of the melting curve of ice~{\sc i} as a function of pressure, a liquid layer could
only exist between ice~{\sc i} and silicates or between ice~{\sc i} and high-pressure ices.
Considering the low surface temperature on OGLE-2005-BLG-390Lb ($T_\mathrm{surf}\ll 250$\,K) and
an adiabatic ocean\footnote{In the following, the term `ocean' is used even when refering to a
layer of liquid water below an ice-{\sc i} shell. In fact, what we describe here is a global
liquid layer that spans within the whole planet, hence it shares similarity with the etymological
origin for `ocean', i.e., the river Okeanos ('$\omega \kappa \varepsilon \alpha \nu \acute{o}
\varsigma$ $\pi o \tau \alpha \mu \acute{o} \varsigma$) that encircles the world in the Greek
mythology.} below ice~{\sc i}, the bottom of the ocean must be at pressure $\la 0.6$~GPa.
Therefore, due to the high gravity on OGLE-2005-BLG-390Lb, the thickness of such an ocean must be
smaller than about 50~km.

\subsection{Heat Flow from the Silicate Mantle}
\label{sec:heatflow} The internal heat flow determines the temperature profile within the planet,
and particularly within the ice layer. Knowing that the most probable age of OGLE-2005-BLG-390Lb
is 10~Gyr, the main source of internal heat at present time is the decay of long-lived radioactive
isotopes of uranium ($^{238}$U and $^{235}$U), thorium ($^{232}$Th), and potassium ($^{40}$K).
Another non-negligible contribution to the heat flow is the secular cooling of the residual heat
produced during the planet formation (20\% for the Earth according to Turcotte \& Schubert 2001).
In the following, we assume that the present heat flow is equal to the radiogenic heat flow.

We take the mass fractions of the radioactive isotopes in the planet when it was 4.56-Gyr old
similar to those of the Earth at the present time. Then, we calculate the heat produced by the
decay of the long-lived isotopes, or radiogenic heat $q_R$. The mass fraction $X(t)$ of a given
species at a given time $t$, in Gyr, is (Turcotte \& Schubert 2001)
\begin{equation}
\label{eq:heat} X(t) = X_{\oplus} \exp{\left[(4.56 - t) \frac{\ln{2}}{\tau_{1/2}}\right]},
\end{equation}
where $X_{\oplus}$ is the terrestrial mass fraction of the species nowadays and $\tau_{1/2}$ is
its half-life. Both are given in Table~\ref{tab:isotopes}. The internal heat production due to the
radiogenic decay, per unit of mass and at a time $t$, is simply $h_\mathrm{rad} = \sum_{i} h_i
X_i(t)$, where $h_i$ is the heat produced by the species $i$ per unit of mass given in
Table~\ref{tab:isotopes}. Finally, the heating per unit of mass of silicate mantle is
$6.2\times10^{-12}$~W\,kg$^{-1}$ at the age of 4.56~Gyr, but it is only
$3.2\times10^{-12}$~W\,kg$^{-1}$ at 10~Gyr, liekly the age of OGLE-2005-BLG-390Lb.

More generally, the heat production is proportional to the abundances of the
radioactive isotopes within the silicates. Shall the planet contain radioactive
material more abundantly than Earth, it would indeed produce more heating. The
effects of a varying heat production are dealt with in
\S~\ref{sec:increased_heating}.

In the following, we use the radiogenic heat produced per unit of mass of the planet,
$h'_\mathrm{rad} = h_\mathrm{rad} \times f_\mathrm{sil}$, where $f_\mathrm{sil}$ is the mass
fraction of silicates in the planet. In fact, the radioactive isotopes are only contained in the
silicate mantle of the planet, not in the metallic core nor in the ice shell. In
\S~\ref{sec:radius}, we saw that $f_\mathrm{sil}$ is estimated to be 68, 51, or 34\%wt, for water
mass fraction of 0.025, 25, or 50\%wt, respectively. At the age of 10~Gyr, this gives
$h'_\mathrm{rad}$ of $2.2\times10^{-12}$, $1.6\times10^{-12}$, and
$1.1\times10^{-12}$~W\,kg$^{-1}$, respectively.

At a depth $z$ within the ice shell and at a time $t$, the heat flow generated by the radioactive
isotopes is
\begin{equation}
\label{eq:Q_R} q_\mathrm{rad}(z,t) = \frac{h'_\mathrm{rad}(t) \times M_P}{4 \pi (R_P - z)^2}.
\end{equation}
At the surface ($z = 0$), for $t=10$~Gyr, and for $M_P$ within [3,11]~\Mearth, we find
$q_\mathrm{rad}$ within [40,70], [24,44], and [13,24]~mW\,m$^{-2}$, in the 0.025, 25, and 50\%wt
cases, respectively.

\section{CAN LIQUID WATER EXISTS WITHIN THE ICE SHELL?}
\label{sec:liquid_now} Although liquid water cannot exist on the surface of the exoplanet
OGLE-2005-BLG-390Lb (\S~\ref{sec:temp}), a liquid layer can exist \emph{below the low-pressure
ice-{\sc i} shell}. This ice shell is heated from below by the decay of radioactive elements. An
adiabatic ocean can be present if the temperature profile crosses the melting curve of ice~{\sc
i}. There are two cases to be considered:

(i) If the planet is water-poor (0.025\%wt), the heat produced by silicates can be transferred
either by conduction or convection across the low-pressure ice-{\sc i} mantle. Examples of
conductive profiles are sketched in Fig.~\ref{fig:camembert_no_h2o}.

(ii) If the planet is water-rich (25 or 50\%wt), the ice mantle consists of two
parts: a low-pressure ice-{\sc i/ii} shell and a high-pressure ice-{\sc vi/vii}
shell (\S~\ref{sec:water-rich shell} and Fig.~\ref{fig:camembert_plenty_h2o}).
The high-pressure mantle allows the heat to be transferred very efficiently by
convection from the silicates up to the base of the low-pressure ice-{\sc i/ii}
mantle, at the interface {\sc ii/vi} (see Fig.~\ref{fig:phase_diagram}). The
heat is then transferred across the low-pressure ice-{\sc i/ii} mantle by
conduction or convection, as in case~(i).

The absence or presence of liquid water can be constrained by considering the
extreme cases when the heat is transferred only by conduction
(\S~\ref{sec:conduction}) or when the convection is set up
(\S~\ref{sec:convection}).

\subsection{Conduction in the Ice}
\label{sec:conduction} In the top layer of ice, the heat is transferred by thermal conduction to
the surface of the planet. The expected temperature profiles are similar to those sketched in
Figs.~\ref{fig:camembert_no_h2o} and \ref{fig:camembert_plenty_h2o} across the low-pressure
ice-{\sc i} and ice-{\sc i/ii} layers, respectively.

The equilibrium between heat flux and heat production rate is assumed, i.e.,
the heat flux at the surface equals the flux heating from below (which
corresponds to the internal heating by silicates, see \S~\ref{sec:heatflow}).
The boundary conditions are: (i) at the surface, the temperature is
$T_\mathrm{surf}=T_\mathrm{surf}(M_P)$ obtained from
eqs.~(\ref{eq:Mplanet_to_Mstar}, \ref{eq:poly_a}, \ref{eq:L_star}, and
\ref{eq:T_eq}), and (ii) at the base of the low-pressure ice layer, the heat
produced $h'_\mathrm{rad}$ is a constant related to the heat flux $q$ through
eq.~(\ref{eq:Q_R}).

The conservation of energy when there is no additional heat source within the ice layer is
\begin{equation} \label{eq:conservation_of_energy}
\frac{\partial q(r)}{\partial r} + \frac{2}{r} q(r) = 0,
\end{equation}
where $q(r)$ is the flux at a radial distance $r$ from the planet center and is given by the
Fourier law in the purely diffusive case,
\begin{equation} \label{eq:fourier_law}
q(r) = -k(T) \frac{\partial T}{\partial r},
\end{equation}
where the thermal conductivity $k$ obeys the empirical law
\begin{equation} \label{eq:conductivity}
k(T) = k_0 / T,
\end{equation}
with $k_0 = 567$~W\,m$^{-1}$ (Klinger 1980).

\begin{figure}
\resizebox{\hsize}{!}{\includegraphics{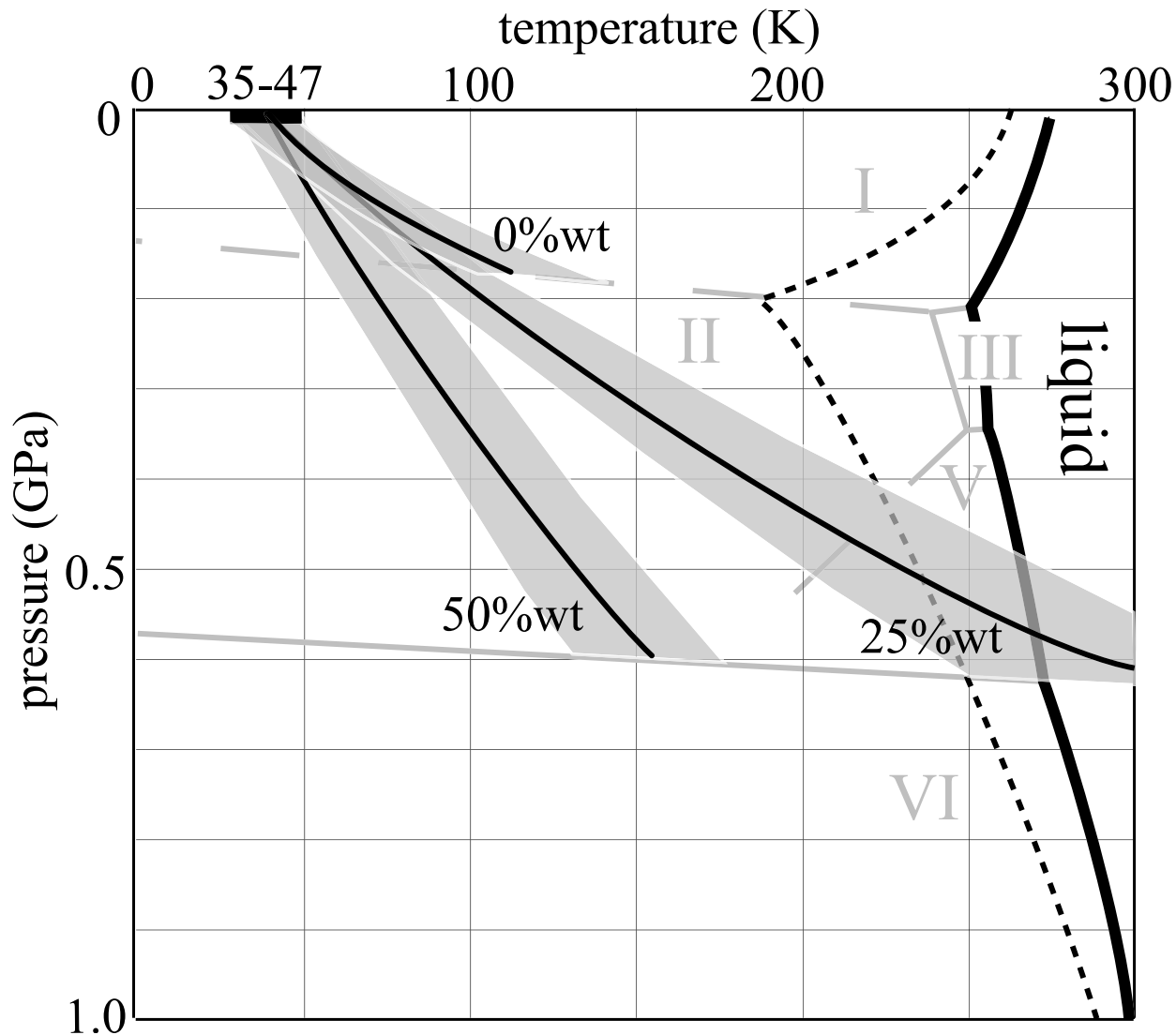}} \caption{Temperature profiles obtained with
the diffusive model in which the heat flow is transferred only through thermal conduction. The
profiles are calculated across the low-pressure ice shell in the 0.025\%wt (i.e., the ice-{\sc i}
layer), 25, and 50\%wt (ice-{\sc i/ii} layer) cases. For each case, the profile is depending on
the planetary mass. The surface temperature is correlated to the planetary mass so that the
surface temperature varies from 35 to 47~K (thick horizontal black line) when the mass varies from
3 to 11~\Mearth. The possible profiles are represented (grey area) for each case. The temperature
profile for the most probable mass (5.5~\Mearth) is evidenced (black line). Thermal conduction is
sufficient to evacuate the heat from below in the 0 and 50\%wt cases, either the ice is pure or
NH$_3$-mixed H$_2$O. Indeed, the temperature profiles do not cross the melting curves (thick and
dashed lines, respectively) for both ice compositions. In the 25\%wt case, the temperature profile
crosses the melting curve. However this is a maximum temperature curve. Convection is likely to
set up and, by efficiently transferring the heat, to avoid the temperature to reach the melting
temperature (see text). \label{fig:phase_diag_diffusion}}
\end{figure}

Combining eqs.~(\ref{eq:Q_R}, \ref{eq:conservation_of_energy}, \ref{eq:fourier_law}, and
\ref{eq:conductivity}), the temperature profile is
\begin{equation}
\label{eq:T(r)} T(r) = T_\mathrm{surf} \exp{\left[\frac{h'_\mathrm{rad} M_P}{4 \pi
k_0}\left(\frac{1}{r} - \frac{1}{R_P}\right)\right]}.
\end{equation}
The resulting profiles are plotted in Fig.~\ref{fig:phase_diag_diffusion} for the 0.025, 25, and
50\%wt cases.

Let $\delta$ be the diffusive coefficient,
\begin{equation}
\delta(r) \equiv \frac{T(r)}{T_\mathrm{surf}}.
\end{equation}
Near the surface, we have $g \approx G M_P/R_P^2$ and then $z \approx p / \rho g$ . Therefore,
replacing $r$ by $R_P - z$ in eq.~(\ref{eq:T(r)}), we get
\begin{equation}
\delta = \delta(p,h'_\mathrm{rad}) \approx \exp{\left(\frac{h'_\mathrm{rad} p}{4 \pi G k_0
\rho}\right)}. \label{eq:delta(p)}
\end{equation}
The coefficient $\delta$ does not depend upon the planetary mass, but only on the pressure $p$ and
the heat flow $h'_\mathrm{rad}$. Knowing the surface temperature, we can plot the temperature
profile over the phase diagram, in Fig.~\ref{fig:phase_diag_diffusion}.

At the pressure of $p_{\textrm{{\sc i}}} = 0.21$~GPa -- the pressure at the triple point {\sc
i/iii}/liquid, see Fig.~\ref{fig:phase_diagram} -- the coefficient $\delta$ is 2.8, 2.2, and 1.7
in the 0.025, 25, and 50\%wt cases, respectively. Resulting temperature ranges are [100,135]~K,
[75,105]~K, and [60,80]~K for the planetary mass within [3,11]~\Mearth.

In the 25 and 50\%wt cases, the ice pressure can reach the pressure of the next phase transition,
$p_{\textrm{{\sc ii}}}=0.61$~Gpa. Then, $\delta$ follows the law
\begin{equation}
\delta(p,h'_\mathrm{rad}) \approx \exp{\left[\frac{h'_\mathrm{rad}}{4 \pi G k_0}
        \left(\frac{p_{\textrm{{\sc i}}}}{\rho_{\textrm{{\sc i}}}}+
              \left(p-p_{\textrm{{\sc i}}}\right) / \rho_{\textrm{{\sc ii}}}\right)\right]},
\label{eq:delta(pii)}
\end{equation}
where $\rho_{\textrm{{\sc i}}}=917$~kg\,m$^{-3}$ and $\rho_{\textrm{{\sc
ii}}}=1\,160$~kg\,m$^{-3}$ are the density of ice~{\sc i} and ice~{\sc ii}. We find $\delta$ is
7.3 and 3.7 for the 25 and 50\%wt cases, respectively. As the heat is transferred by thermal
conduction, by using the approximation $T(p_{\textrm{{\sc ii}}})\approx \delta(p_{\textrm{{\sc
ii}}}) T_\mathrm{surf}$, we get temperatures within [250,340]~K and [130,175]~K in the 25 and
50\%wt cases, respectively.

\subsection{Convection in the Ice}
\label{sec:convection} Unlike conduction, convection is an efficient heat
transfer mechanism. If convection is possible in the solid ice shell of
OGLE-2005-BLG-390Lb, then the temperature profile can be described as, starting
from the surface: a thin conductive lid of ice, an upper thermal boundary layer
(TBL), a well-mixed adiabatic ice layer where the temperature $T_\mathrm{conv}$
is nearly constant, and a lower thermal boundary layer (Hussmann et al.\ 2002;
Spohn \& Schubert 2003; Sotin \& Tobie 2004).

According to Spohn \& Schubert (2003), the thermal flux through the convective ice layer is
\begin{equation}
\label{eq:flux conv} q_\mathrm{conv} = b \left(\frac{\alpha
g}{\kappa\eta_\mathrm{conv}}\right)^{\beta} k\Delta T^{1+\beta} \Delta z_\mathrm{conv}^{3\beta-1},
\end{equation}
where the thermal diffusivity $\kappa = 1.47 \times 10^{-6}$~m$^2$\,s$^{-1}$ and the thermal
expansion coefficient $\alpha = 1.56 \times 10^{-4}$~K$^{-1}$ stand for ice~{\sc i}, and
$\eta_\mathrm{conv}$ is the viscosity of the convective layer of thickness $\Delta
z_\mathrm{conv}$. The viscosity $\eta$ depends on temperature as
\begin{equation}
\label{eq:viscosity} \eta(T) = \eta_0 \exp\left[27\left(\frac{T_m}{T} - 1 \right)\right],
\end{equation}
with $\eta_0 = 5 \times 10^{13}$~Pa\,s, the viscosity at the melting temperature $T_m$. The
temperature difference $\Delta T$ across the convective layer is twice the temperature difference
across the two thermal boundary layers ($\Delta T_\mathrm{TBL}$) given by (Sotin \& Tobie 2004)
\begin{equation}
\label{eq:DeltaT tbl} \Delta T_\mathrm{TBL} = T_m - T_\mathrm{conv} = -1.43 \times
\left[\left.\frac{\partial \ln(\eta)}{\partial T}\right|_{T=T_\mathrm{conv}}\right]^{-1},
\end{equation}
where $T_m$ and $T_\mathrm{conv}$ are the melting temperature and the temperature in the
convective well-mixed adiabatic ice layer.

Spohn \& Schubert (2003) use typical values  $b = 0.2$ and $\beta = 0.25$, while Deschamps \&
Sotin (2001) use $b=0.79$ and $\beta=0.263$. Large $b$-values correspond to very efficient heat
transport by convection. Then, the temperature remains close to the surface temperature, and
hardly reaches the melting temperature. The presence of ice is favored. On the contrary, low
$b$-values favors the presence of an ocean below a convective ice layer. In the following, we will
consider both sets of values for $b$ and $\beta$.

\subsection{Constraints on the Presence of Liquid Water}

\subsubsection{Method}

To have a liquid layer below the ice shell, the temperature must reach the melting temperature
above the triple point {\sc i/iii}/liquid at 0.21~GPa and 251~K; in other words, the temperature
gradient below the surface must be steep enough to reach the melting temperature within the
ice-{\sc i} shell. Then, below this level, we have an ocean whose depth is limited by the pressure
above which the liquid water is compressed into high-pressure ice (Fig.~\ref{fig:camembert_no_h2o}
and~\ref{fig:camembert_plenty_h2o}).

Using the equations of \S~\ref{sec:conduction} and \ref{sec:convection}, we can evaluate the
minimum pressure -- when only conduction drives the heat flow -- and the maximum pressure -- when
there is a convective layer below a conductive lid -- at which the temperature reaches the melting
temperature. If the minimum pressure is larger than the triple point pressure, no ocean is
possible below the ice shell. If the maximum pressure is smaller than the triple point pressure
(i.e., if the temperature reaches the melting temperature despite the very efficient heat transfer
by a convective layer), then the presence of an ocean is warrantied.

\subsubsection{Results}

Taking the temperature profile obtained when only conduction drives the heat flow
(Fig.~\ref{fig:phase_diag_diffusion}), we see that it does not reach the melting temperature in
the 0.025 and 50\%wt cases. Therefore, an ocean within the ice-{\sc i} shell of
OGLE-2005-BLG-390Lb cannot be present. In the 0.025\%wt case, there is simply not enough water and
the maximum temperature reached at the bottom of the ice shell is in any cases well below 150~K.
Thus, no liquid water is possible. In the 50\%wt case, deeper in the planet, the maximum
temperature reached at the bottom of the ice-{\sc ii} shell is in any cases well below 250~K. This
does not allow for the presence of liquid water either.

Only in the 25\%wt case, the temperature profile crosses the melting curve, although it does so
below the triple point {\sc i/iii}/liquid, at pressure of about 0.5~GPa. This means that
conduction alone is not able to transfer the heat from below across the solid ice shell. A
conductive equilibrium does not exist in this case and the temperature profile plotted in
Fig.~\ref{fig:phase_diag_diffusion} is not realistic.

Convection is likely to set up in the high-pressure ice and the temperature gradient can be
steeper than the one calculated with conduction only. Using eq.~(\ref{eq:flux conv}) for the heat
flow at 10~Gyr and other equations of \S~\ref{sec:convection}, and taking the parameters $b$ and
$\beta$ typical of a very efficient heat transport by convection (Deschamps \& Sotin 2001), we
find that the convective layer can be thick enough to allow the heat to be transferred without the
temperature reaching the melting temperature. The presence of liquid water is not allowed in that
case. However, if less efficient convection is considered by using smaller $b$-values (Spohn \&
Schubert 2003), we find that the temperature can effectively reach the melting temperature within
the ice-{\sc ii} layer. In this very last case, solid-state convection cannot evacuate the heat
from below, therefore the ice-{\sc ii/v} shell must be partly melted.

Nevertheless, in all the cases considered, the temperature profile never crosses the melting curve
above $p_{\textrm{{\sc i}}} = 0.21$~GPa. Therefore, an ocean cannot exist between ice~{\sc i} and
higher-pressure ices.

\subsubsection{Conclusion}

Our study in which (i) the surface temperature is as cold as the equilibrium temperature, (ii) the
ice is pure H$_2$O-ice, and (iii) the internal heat only consists in the radiogenic heat, implies
that the ice-{\sc i} shell of OGLE-2005-BLG-390Lb must be entirely frozen whatever the amount of
water and the mass of the planet. Thus, no ocean can be present between ice~{\sc i} and
high-pressure ice.

Only in the case of an intermediate amount of water, around 25\%wt, there is a possibility to have
liquid water within the ice-{\sc ii/v} shell because both thermal conduction and low-efficient
solid-state convection are unable to remove the internal heat of the planet. If, however,
solid-state convection is as efficient as proposed by Deschamps \& Sotin (2001), then no liquid
water exist in the ice-{\sc ii/v} layer of the 25\%wt case.

In fact, the existence of liquid water is the result of a competition between two fundamental
parameters: the heating $h'_\mathrm{rad}$ on one hand and the thickness of the ice layer on the
other hand. Given the mass of the planet, the more water the planet contains, the less massive the
silicate mantle is, and thus the more water in the planet, the less radiogenic heat there is to
warm the ice. Meanwhile, the more water in the planet, the thicker the ice shell can be.
Therefore, higher temperatures can be reached in the ice shell above the silicate mantle. This
competition explains why the intermediate case with 25\%wt water appears to be the most favorable
for the presence of liquid water in OGLE-2005-BLG-390Lb, while in both 0.025 and 50\%wt cases,
there are no possibilities for the presence of liquid water.

\section{VARYING PARAMETERS}
\label{sec:freeparameters}

In the previous section, we found that if the surface temperature of OGLE-2005-BLG-390Lb is equal
to the equilibrium temperature (from 35 to 47~K for planetary masses of 3 to 11~\Mearth,
respectively) and the ice composition is 100\% H$_2$O, then no oceans can be found between
ice-{\sc i} and high-pressure ices. We will discuss now the effects of a surface temperature
larger than the equilibrium temperature, of a different composition of the ice, and finally of a
larger internal heating.

\subsection{Effect of an Increase in Surface Temperature}
\label{sec:greenhouse} The possibility of the presence of an atmosphere thick enough to trigger a
greenhouse effect or a smaller albedo could rise significantly the surface temperature. For
instance, an albedo close to zero would increase the equilibrium temperature by a factor
$\sqrt[4]{2}$, that is about 19\%. Therefore, we shall evaluate the minimum increase in the
surface temperature that allows the existence of an ocean between low- and high-pressure ices.

As a first approach, we can calculate the maximum temperature at the pressure $p_{\textrm{{\sc
i}}}$ of the triple point {\sc i/iii}/liquid with thermal conduction only
(\S~\ref{sec:conduction}). On one hand, for a given heat flow and at the pressure of the triple
point corresponding to the base of the low-pressure ice shell, there is a diffusive coefficient
$\delta(\mathrm{p_{\textrm{\sc i}}})$. On the other hand, the ratio of the melting temperature at
the triple point ($T_{m\textrm{\sc i}}=251$~K) to the actual surface temperature
($T_\mathrm{surf}$) is equivalent to a minimum diffusive coefficient above which the water can
melt. An ocean could exist if the increase of the surface temperature is above $\Delta
T_\mathrm{surf}$ where
\begin{equation}
\delta(p_{\textrm{{\sc i}}}) \times ( T_\mathrm{surf}+\Delta T_\mathrm{surf}) = T_{m\textrm{{\sc
i}}}.
\end{equation}

Using this equation, we find that an increase of the surface temperature of at least [40,55],
[65,80], and [100,115]~K, for the 0.025, 25, and 50\%wt cases, respectively, are required in order
to partly melt the ice layer. However, those values cannot be obtained with smaller albedo and are
unlikely to result from a greenhouse effect, given the small energy the planet is receiving from
its star and the tenuous atmosphere of N$_2$.

\subsection{Effect of Ice Composition}
\label{sec:ammoniac} We have assumed that the ice shell of OGLE-2005-BLG-390Lb is only made of the
different phases of H$_2$O. However, due to the low temperature at the surface of the planet (from
35 to 47~K, see \S~\ref{sec:temp}), not only H$_2$O, but also NH$_3$, CH$_4$, and possibly N$_2$
can condense. An accurate model of the ice shell would therefore need including these species in
addition to H$_2$O, but is beyond the scope of this paper. A number of detailed studies have shown
that a mixture of H$_2$O and, e.g., NH$_3$ crystallizes at a lower temperature than pure H$_2$O
(see, e.g., Sotin et al.\ 1998; Spohn \& Schubert 2003). The melting curve of this mixture is
plotted in Figs.~\ref{fig:phase_diagram} and \ref{fig:phase_diag_diffusion}.

It shows that our previous results remain unchanged. In fact, the maximum temperatures calculated
at the base of the ice-{\sc i} shell are below the crystallizing temperature of both ammoniated
and pure water in all cases. Therefore, no ocean is possible below the ice-{\sc i} lid, even with
the presence of NH$_3$.

\subsection{Effect of an Increase in the Internal Heating}
\label{sec:increased_heating} Similarly to the estimation of the surface temperature in
\S~\ref{sec:greenhouse}, we can calculate how much the internal heating has to be increased in
order to partially melt the frozen ice shells. The minimum heat production $h'_\mathrm{liq}$ above
which the temperature can reach the melting temperature $T_{m\textrm{\sc i}}$ at pressure of the
triple point {\sc i/iii}/liquid, $p_{\textrm{{\sc i}}}$, can be estimated using results of
\S~\ref{sec:conduction} and is given by:
\begin{equation}
h'_\mathrm{liq} = h'_\mathrm{rad} \times \frac{\ln \left[T_{m\textrm{\sc
i}}/T_\mathrm{surf}(M_P)\right]}
     {\ln\left[\delta(p_{\textrm{{\sc i}}},h'_\mathrm{rad})\right]}.
\end{equation}
We find that the heat production calculated at 10~Gyr has to be multiplied by a factor of at least
1.6 to 1.9 in order to have liquid water in the 0\%wt H$_2$O case, a factor of 2.1 to 2.5 in the
25\%wt H$_2$O case, and 3.2 to 3.8 in the 50\%wt H$_2$O case.

If we allow for convection as in \S~\ref{sec:convection}, we obtain larger heat production rate.
If we use small values of $b$ corresponding to a convection inefficient for the transport of the
internal heat, the results are barely changed. If we assume a much more efficient convection in
the ice-{\sc i} shell, we find that the heating rates needed to obtain liquid water are 75 to
100\% larger than those obtained by neglecting the convection. For instance, for a 5.5-\Mearth\
planet with 25\%wt H$_2$O, the heating rate needed to reach the melting temperature within the
low-pressure ice is 3.9 and 2.2 higher than the heating at 10~Gyr, for a large and a small $b$,
respectively. Those results are barely changed by the possible presence of NH$_3$.

Although the present-day OGLE-2005-BLG-390Lb is likely to be entirely frozen,
the heating needed to have liquid water below the surface of this planet is
only a factor-of-a-few larger than the calculated radiogenic heating at 10~Gyr.
It must be recalled that this radiogenic heating is only a minimum: a remnant
heat from the accretion is certainly not negligible, in particular for such a
massive planet. A comparison with the Earth shows that it could be of the same
order as the radiogenic heating. Moreover, in the past the radiogenic heat was
larger. Using eq.~(\ref{eq:heat}), the radiogenic heat is a factor of 2 and 4
larger when the planet was 4.4- and 1.8-Gyr old. This allows us to suggest that
liquid water was certainly present in the past during several billion years
below the surface of OGLE-2005-BLG-390Lb.

\section{CONCLUSIONS}

\label{sec:conclusion} Because OGLE-2005-BLG-390Lb orbits a few \AU\ away from a faint M star and
is about 10-Gyr old, this planet is likely to be entirely frozen.

In the favored scenario, the planet has no heavy atmosphere and its surface temperature is close
to the equilibrium temperature of a black body (35 to 47~K); then, the surface should be entirely
frozen.

If the planet contains terrestrial amounts of water (or much more), a $\la 10$-km- (or $\ga
1\,000$-km-) thick ice mantle can be present. The radiogenic heat production rate, which depends
on the quantity of water/silicates contained in the planet, is probably not sufficient to prevent
a potential sub-surface ocean from freezing completely at the likely age of 10~Gyr.

However, remnant heat and larger radiogenic heating in the past allow us to suggest that a liquid
ocean must have been present during several billion years. In similarity with the icy satellites
of the Solar system, this liquid water was located below low-pressure ice~{\sc i}. At the bottom
of the ocean, either a silicate mantle or high-pressure ices could be found, for water-poor or
water-rich planet, respectively.

In some respects, OGLE-2005-BLG-390Lb opens a new frontier in modeling extra-solar planets: not
only because it is presently the closest planet to the Earth in terms of mass, but also because it
is located at the cold border of its planetary system, beyond the snowline. Up to now, the
greatest interest has been shown in observing and modeling hot Jupiters and planets close to their
star, mainly because they were the first to be detected. Microlensing detection technique has now
the potential to unveil a population of ice giants, frozen ocean-planets, and other snowball
Earths. The present work is a first approach to such objects. It opens new perspectives for
exoplanetology.

\acknowledgements We warmly thank Christophe Sotin and Alain L\'eger for useful
advices and comments about the manuscript, Roger Ferlet for thorough reading
and wise suggestions, and the referee Hauke Hussmann for a constructive review.


\clearpage

\begin{deluxetable}{lcc}
    \tablecaption{Parameters of the microlensing event OGLE-2005-BLG-390.\tablenotemark{a}\label{tab:lens}}
    \tablehead{\colhead{Parameter} & \colhead{Symbol} & \colhead{Value}}
    \startdata
        planet-to-star mass ratio & $q$             & $(7.6\pm0.7) \times 10^{-5}$     \\
        projected separation      & $d$             & $1.609\pm0.004~R_\mathrm{E}$    \\
        distance to the source    & $D_\mathrm{S}$  & $8.5$~kpc                       \\
        distance to the lens      & $D_\mathrm{L}$  & $7.2\pm0.8$~kpc                 \\
        mass of the star\tablenotemark{b}      & $M_\star$       & $0.22^{+0.21}_{-0.11}$~\Msun    \\
        mass of the planet\tablenotemark{b}    & $M_P$  & $5.5^{+5.5}_{-2.7}$~\Mearth     \\
        semi-major axis\tablenotemark{b}       & $a$             & $2.6^{+1.5}_{-0.6}$~\AU         \\
    \enddata
    \tablenotetext{a}{From Beaulieu et al.\ (2006).}
    \tablenotetext{b}{The quantity is the mean value of the probability distribution.}
\end{deluxetable}

\clearpage

\begin{deluxetable}{cccc}
    \tablecaption{Properties of long-lived radioactive isotopes in the Earth's interior. \label{tab:isotopes}}
    \tablehead{
        \colhead{Isotope} & \colhead{Heat Production}   & \colhead{Half-Life}          & \colhead{Concentration} \\
        \colhead{}        & \colhead{$h$ (W kg$^{-1}$)} & \colhead{$\tau_{1/2}$ (Gyr)} & \colhead{$X$ (kg kg$^{-1}$)}}
    \startdata
        $^{238}$U & $9.37\times10^{-5}$   & 4.47                    & $25.5\times10^{-9}$        \\
        $^{235}$U & $5.69\times10^{-4}$   & 0.704                   & $1.85\times10^{-10}$       \\
        $^{232}$Th& $2.69\times10^{-5}$   & 14.0                    & $1.03\times10^{-7}$        \\
        $^{40}$K  & $2.79\times10^{-5}$   & 1.25                    & $3.29\times10^{-8}$        \\
    \enddata
    \tablecomments{The concentration given are the value in the Earth mantle.
        From Cox (2000), referring to Turcotte \& Schubert (1982).}
\end{deluxetable}


\clearpage

\clearpage

\clearpage

\clearpage

\clearpage

\clearpage

\end{document}